\documentclass[10pt,twocolumn]{wlscirep}

\usepackage{mathpazo}
\usepackage[tt=false]{libertine}
\usepackage{helvet}

\DeclareCaptionLabelFormat{supportinginfo}{#1~#2}

\def\myappendix{\appendix%
\renewcommand\thefigure{\thesection.\arabic{figure}}
\renewcommand\thetable{\thesection.\arabic{table}}
\setcounter{figure}{0}
\renewcommand\thetable{\thesection.\arabic{table}}
\setcounter{table}{0}
\renewcommand{\theequation}{\thesection.\arabic{equation}}
\setcounter{equation}{0}
\captionsetup{labelformat=supportinginfo}
}

\graphicspath{{img/}}

\usepackage{siunitx}
\usepackage{mathtools}

\usepackage{booktabs}
\usepackage{tabularx}
\usepackage{adjustbox}

\usepackage{rotating}

\usepackage{microtype}

\usepackage{algorithm}
\usepackage{algpseudocode}

\usepackage{listings}
\usepackage{courier}
\usepackage{xcolor}

\definecolor{codegreen}{rgb}{0,0.5,0}    
\definecolor{codepurple}{rgb}{0.5,0,0.5} 
\definecolor{codegray}{gray}{0.5}        

\lstdefinestyle{codestyle}{
    columns=fullflexible,
    basicstyle=\ttfamily\footnotesize,
    keywordstyle=\bfseries\color{blue},       
    commentstyle=\itshape\color{codegreen},   
    stringstyle=\color{codepurple},           
    numberstyle=\tiny\color{codegray},        
    emph={%
CircularAperture,KtZ,XTauKtZ,XNew,PhaseRamp,XTauZ,Equation5,Fz,Z,X,
poisson_norm,sum_squares,elemwise_multiply,crop,fft_inv,bayer,Variable,%
split,group,absorb,scale},           
    emphstyle=\color{orange}\bfseries,        
    breaklines=true,
    captionpos=b,
    keepspaces=true,
    numbers=left,
    numbersep=5pt,
    showspaces=false,
    showstringspaces=false,
    showtabs=false,
    tabsize=2
}

\title{Designing across domains with declarative thinking: Insights from the 96-Eyes ptychographic imager project
}

\author[1,$\ast$]{Antony C.~Chan}

\affil[1]{Consultant, high-throughput microscopy and hardware-accelerated algorithms,
San Diego, CA 92131, USA}
\affil[$\ast$]{Correspondence: \url{https://www.linkedin.com/in/antonycschan/}}


\begin{abstract}
This article presents a practitioner's reflection on applying declarative, 5th generation, problem formulation language (5GL) to \emph{de novo} imaging system design,
informed by experiences across the interdisciplinary research in academia and cross-functional product development within the private sector.
Using the 96-Eyes project: 96-camera parallel multi-modal imager for high-throughput drug discovery as a representative case,
I illustrate how project requirements, ranging from hardware constraints to life sciences needs, 
can be formalized into machine-readable problem statements to preserve mission-critical input from diverse domain stakeholders.
This declarative approach enhances transparency, ensures design traceability, and minimizes costly misalignment across
optical, algorithmic, hardware-accelerated compute, and life sciences teams.

Alongside the technical discussion of 5GL with real-world code examples,
I reflect on the practical barriers to adopting 5GL in environments where imperative,
3rd-generation languages (3GL) remain the default medium for inter-team collaboration.
Rather than offering an one-size-fits-all solution,
these learned lessons highlight how programming paradigms implicitly shapes research workflows through existing domain hierarchies.
The discussion aims to invite further explorations into how declarative problem formulations can
facilitate innovation in settings where concurrent R\&{}D workflows are gaining traction,
as opposed to environments where sequential, phase-driven workflows remain the norm.

\end{abstract}

\begin{document}
\def\um{\,\si{\micro\meter}}
\def\mm{\,\si{\milli\meter}}
\def\nm{\,\si{\nano\meter}}
\def\Hz{\,\si{\hertz}}

\maketitle
\thispagestyle{fancy}

\begin{figure*}[tb]
\centering\includegraphics[width=0.7\textwidth]{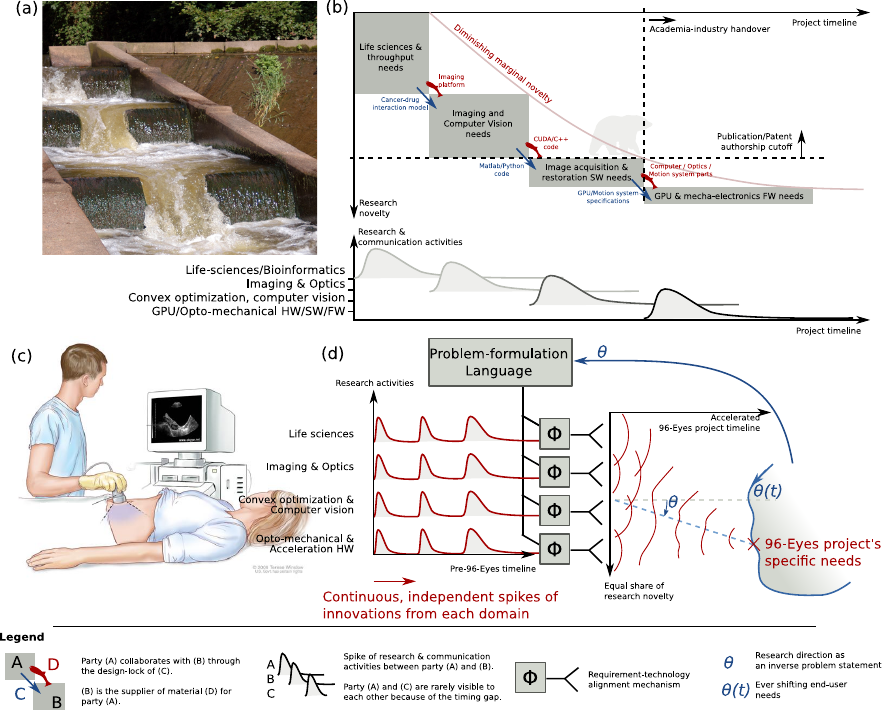}
\caption{\label{fig:waterfall}%
{\bf Distinctions between waterfall-style \emph{research collaboration} and concurrent research \emph{strategic initiatives}.}
(a)~Waterfall-style R\&{}D is analogous to a fish climbing a ladder during peak salmon season.
(b)~Waterfall-style exhibits intermittent spikes of research activities, followed by successive hand off of the project requirements with diminishing research novelty returns on investment. This style tightly couples the timing of research domain ``focus'' and the timing of inter-domain communications,
resulting in spikes of talent acquisition activities and the risks of the distortion/loss of institutional knowledge.
(c)~Concurrent R\&{}D is analogous to Doppler ultrasonic imaging with a phase-array probe, which electronically aligns the individual pulse wavelets to focus on the target tissue.
(d)~Concurrent R\&{}D is most effective when the end-user needs are accurately described as the \emph{inverse problem statement} through the problem formulation language, which in turn is disseminated over domain silos to ensure equal research contribution and expectation alignment.
The original design intent is preserved in DSL, and the institutional knowledge is preserved through publications on the corresponding freestanding, domain-specific research journals.
Acronyms: hardware (HW); software (SW); firmware (FW); artificial intelligence (AI); graphical processing units (GPU); original equipment manufacturer (OEM).
Photo courtesy of: (a)~Wikipedia; (c)~Terese Winslow and the scientific education article on the NIH website.}
\end{figure*}

\section{Motivation: the complexity of 96-Eyes project justifies concurrent R\&{}D approach}

In the Apollo~13 mission, space capsule damage stranded the astronauts on the way to the moon. Back on Earth, there were heroic efforts to save the astronauts from CO2 poisoning by retrofitting a round-shaped ventilator port with a squared-shaped CO2 scrubber cartridge amid limited resources aboard.
The ``successful failure'' was universally praised as an effective teamwork in a fast-paced, high-stake environment.
The legend was later dramatized in Hollywood blockbusters.
However, as the adrenaline level subsided, one begs the question of whether the \emph{impedance mismatch} crossing domain boundaries --- having to fit a square peg through a round hole --- is necessary in the first place.

The author draws inspirations from the prediction of the so-called multidiscipline joint design ``utopia'', pioneered by the model-based system engineering community~\cite{Grady2008}.
As illustrated at the bottom of Fig.~\ref{fig:waterfall}(d), the ultimate goal of a research strategic initiative is to map the overarching needs to the  technologies well known by the respective domains through continuous innovation.
The set of domain-specific work can be independently sponsored by the corresponding agencies, and yet through the use of a common description of the problem statement, the interests are aligned to produce a working solution.
The research outputs from the project in turn results in \emph{freestanding, independent journal publications} of the corresponding research domains.

In practice, such a joint design utopia is rarely seen in academic and industries. Most of the research collaboration (within academia or between academia and industries) takes the form of \emph{waterfall style}\cite{Kedia2022} research approach [Fig.~\ref{fig:waterfall}(a)].
Under the backdrop of intense competitions and the promise to deliver timely results, project stakeholders put a higher focus on addressing the top domain needs having the highest impact, followed by the next domain in descending order of research novelty.
To goal is to apply the Pareto's principle\cite{Newman2005} to the research effort; roughly 80\% of the research novelty should be demonstrated in the top 20\% of the R\&{}D domains contributing to the project.
As illustrated in Fig.~\ref{fig:waterfall}(b), one dominating domain (e.g.~determining the drug-cell line interaction model to address the high-throughput drug screening needs) takes the blunt of heavy lifting to de-risk the project. The research output is then shared to downstream R\&{}D domains through various design-locked documents,
such as cell line and drug interaction models, wet-lab protocols, algorithm source code, and electronic/mechanical schematics.
Such a \emph{sequential R\&{}D} approach introduces a tight coupling of the timing of the technology transfer, and the timing of the domain-specific research activities.
This is as if the knowledge (e.g.~optical imaging techniques) and constraints (e.g.~optics performance specifications) are ``flowing'' one-way down the metaphorical ``waterfall'', and the implementation (e.g.~instrument control code, opto-mechanical schematics) swimming upstream against the flow, like a salmon.

\section{Case study: Multi-modality, parallel, and lateral motion free microscopy for 96-well plates requires unconventional FPM modelling and HW-acceleration}

\begin{figure}[tb]
\centering
\includegraphics[width=\columnwidth]{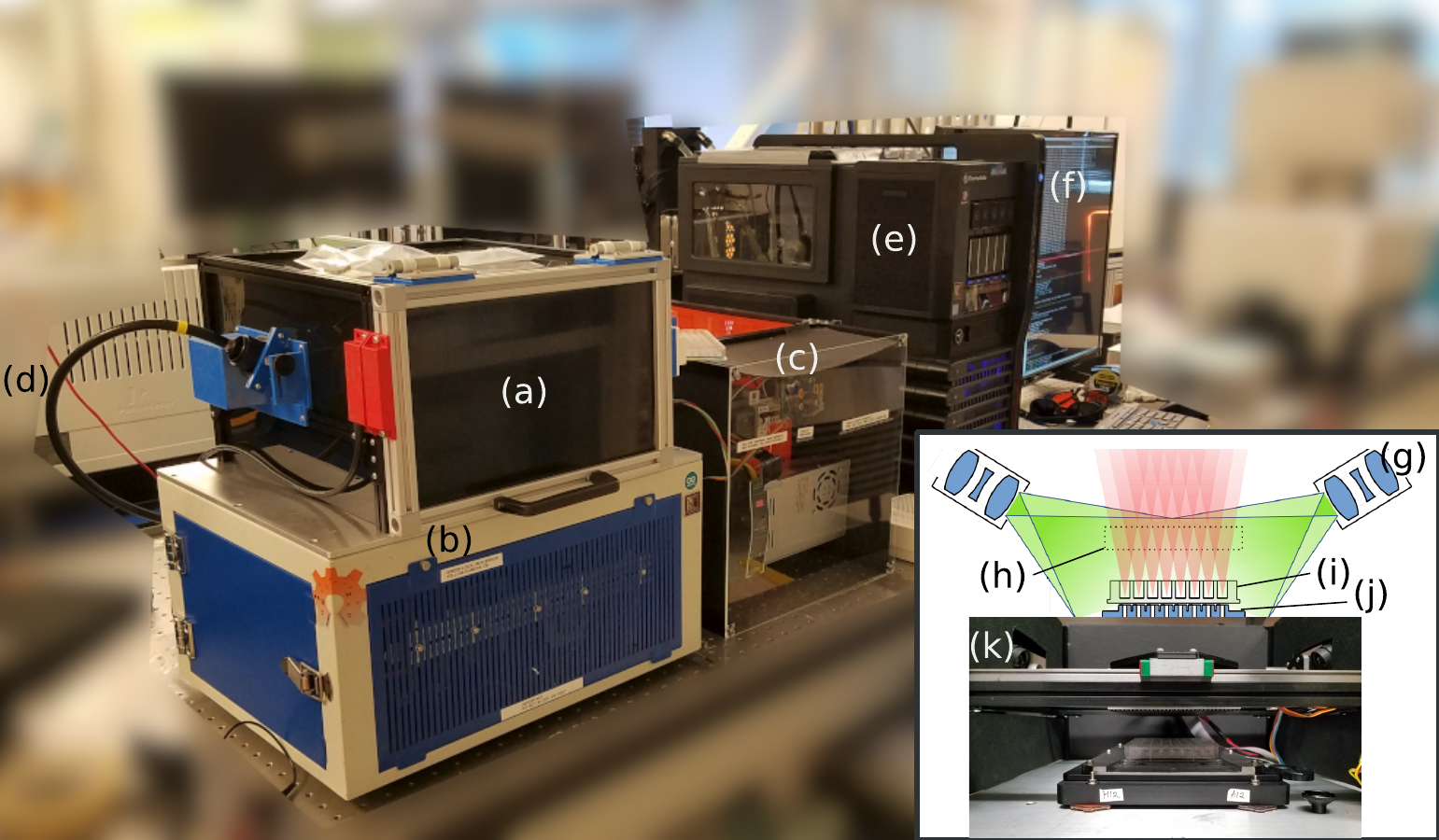}
\caption{\label{fig:setup}%
{\bf The 96-eyes intrument's most recent design iteration.}
The hardware consists of
(a)~plate cradle and incubation module;
(b)~96-camera and power distribution module;
(c)~external illuminator as the dual-wavelength laser sources ($465\nm$ and $520\nm$),
transmitting excitation lights through (d)~a pair of homogenizing light pipes;
(e)~multi-GPU compute workstation; and
(f)~the 49-inch display.
(Inset)~The front view of the plate cradle and the illustration.
(g)~Projection lens pair at the distal end of the homogenizing light pipes;
(h)~LED matrix illuminator;
(i)~96-well plate;
(j)~96-microscope objective array.
The new design also incorporates (k)~a customized piezo-flexure wide aperture stage for z-stack fluorescence imaging.
Computational phase imaging remains completely motion-free thanks to the Ptychographic extended depth of focus feature.}
\end{figure}

The level of sophistication of the 96-Eyes project justifies the adoption of the concurrent R\&{}D over waterfall-style.
A quick recap: the 96-eyes imaging system is a parallel motion-free microscopy system for high-throughput screening, capable of simultaneously imaging all wells on the 96-well cell culture plate. Since its first inception featuring $xyz$-motion-free computational refocusing of stain-free cell cultures\cite{Chan2019}, the 96-eyes instrument underwent an upgrade to incorporate an additional fluorescence channel (excitations $= 465\nm,\ 520\nm$, emission $= 510\nm,\ 625\nm$), as well as the ability to conduct an one-off z-stack fluorescence image capture step per plate (Figs.~\ref{fig:setup}) with a wide-aperture z-axis stage.

Compared to similar optical system architectures\cite{Zheng2013,Kim2016,Tian2015}, the 96-Eyes is unique in the multifaceted needs:
 the life science needs for the walkaway automation,
 the data throughput imaging needs for parallel web-lab assays, and
 the engineering cost control needs for the low-cost imaging optics with consumer-grade imaging sensor arrays.
Mutual, effective multidisciplinary dialogues and technology cross-pollinations among researchers from diffraction optics, image formation theory, computer vision, and compute hardware are crucial for the timely delivery of the project.
The 96-Eyes serves as an ideal case study because the projects aims to bridge between
the pull from \emph{user needs}: high-throughput drug screening, and the
\emph{technology push}: Fourier Ptychographic microscopy setup\cite{Zheng2013} and the robust phase retrieval algorithm\cite{Ou2014}.
The work scope and the engineering challenges the project poses can be overwhelming;
one cannot simply (metaphorically) throw the requirement documentations
(e.g. research articles, wet-lab assay protocols, Matlab code, optical breadboard) down the waterfall and wish the downstream domain experts good luck.
As the domain-specific, tribal knowledge is disseminated (and likely aberrated) downstream through successive design-locks, a mere misunderstanding of the materials cause the project to derail.
A new system-level design workflow needs to be sought.

\section{Inverse problem statement as a declarative coding language}

I argue that the underlying reason for the work friction in the waterfall-style R\&{}D pipeline stems from the use of low-level code (e.g. Matlab, Python),
aka 3th generation language (3GL, Table~\ref{tab:language-generation}), as the primary medium of communication among domain experts. To elaborate, we can dissect an excerpt of the FPM phase retrieval algorithm,
which aims to recover the thickness distribution of the cell line (aka phase image) with the alternating projection algorithm in Alg.~\ref{alg:fpm-code}.

\def\Restored{\hat u}
\def\Gain{\eta}
\def\Bayer{\mathbf{M}}
\def\Fourier{\mathbf{F}}
\def\Blur{\mathbf{H}}
\def\ObliqueIlluminate{\mathbf{Q}_j}
\def\Raw{\mathbf{b}_j}
\def\FlatField{\mathbf{1}}

\def\ImagePlane{v_j^{k+1/2}}
\def\NewImagePlane{v_j^{k+1}}

\begin{algorithm}
\caption{Pseudocode of the phase retrieval algorithm for 96~Eyes instrument, excluding computational lens aberration compensation and auto-focusing.\label{alg:fpm-code}}

\begin{algorithmic}[1]
\State \textbf{Input:} $I_j$ low resolution image illuminated by the $j$-th oblique light source.

\State \textbf{Initialize}
Wavefront in the camera pupil plane $\tilde u=\Fourier \sqrt{I_1}$.

\For{$k = 1 \ldots V$}
	\For{$j = 1 \ldots 25$}
	    \State $\ImagePlane = \Fourier^T \Blur \ObliqueIlluminate \tilde u$.
	    \Comment Forward propagation

	    \State $\NewImagePlane = \sqrt{I_j} \frac{\ImagePlane}{| \ImagePlane |}$
	    \Comment Image plane update
\label{enum:object-update}

		\State $\Delta u_j^{k+1} = \ObliqueIlluminate^T \overline{\Blur} \Fourier ( \NewImagePlane - \ImagePlane )$
		\Comment Backward propagation

		\State $\tilde u \gets \tilde u + \alpha \Delta u_j^{k+1}$
		\Comment Pupil plane update
    \EndFor
\EndFor
\State \Return{$u$}
\end{algorithmic}

\end{algorithm}

Here, to cater to the general audience, we will only focus on the semantic meaning of the expressions, not the intimidating Maths and symbols.
In plain English: ``Algorithm~\ref{alg:fpm-code} simulates the low-resolution image by \emph{illuminating} the object with oblique illumination, then \emph{forward propagating} the wavefront from the object plane to the image plane through the imaging optics.
Next, the intensity component is \emph{replaced} with the square root raw intensities.
The updated wavefront then \emph{undergoes backward propagation} to the object plane to \emph{update} the earlier object estimate.
And \emph{repeat} until the maximum number of iterations is reached.''

Note the \emph{imperative} language pattern in the algorithm, i.e.~Do-(something), Do-not-do-(something).
It dictates \emph{how} the data should be scaled, transformed, and moved from one state and another state,
but it does not describe \emph{what} kind of problem is being solved under \emph{what} assumptions.
Furthermore, it does not convey \emph{why} there is a problem to solve in the first place.

\begin{table*}
\caption{Curated programming languages used in the 96-Eyes project, grouped by the language generation.
The sixth generation language (6GL) and its role in the 96-Eyes project will be discussed in a separate article.
Abbreviations: 1GL -- 1st generation language;
SW -- software;
HW -- hardware;
R/W -- (disk) read or write;
CAD -- computer aided design;
CAM -- computer aided manufacturing;
API -- application programming interface;
I2C -- inter-integrated circuit protocol.\label{tab:language-generation}}

\resizebox{\textwidth}{!}{%
\begin{tabular}{cccc}
\toprule
    & Description                   & Examples                & Application in the 96-Eyes project \\
\midrule
6GL   & Diagramming \& networking language         & PlantUML, MBSE-lite\cite{Chan2020}   & End-to-end HW/SW requirement traceability analysis\cite{Kass2016} \\
5GL   & Problem formulation language  & ProxImaL\cite{Heide2016}              & Image restoration algorithm generation \\
4GL   & CAD-CAM language              & Halide\cite{RaganKelley2012}, SQL           & Algorithm GPU acceleration; database management\\
3.5GL & Fluent API\cite{Fowler2011} & Taskflow\cite{Huang2022}, CADQuery\cite{Machado2019}    & Concurrent GPU \& disk R/W; mechanical design\\
3GL   & General purpose language      & Python, C++           & Opto-mechanical control automation\\
2GL   & Assembly language             & I2C, NvPTX, SCPI\cite{Bode1999}      & GPU warm up; Stage motion commands \\
1GL   & Transistor-transistor logic   & LED matrix commands    & Programmable oblique illumination for FPM\\
\bottomrule
\end{tabular}%
}
\end{table*}

\begin{figure}
\centering\includegraphics[width=\columnwidth]{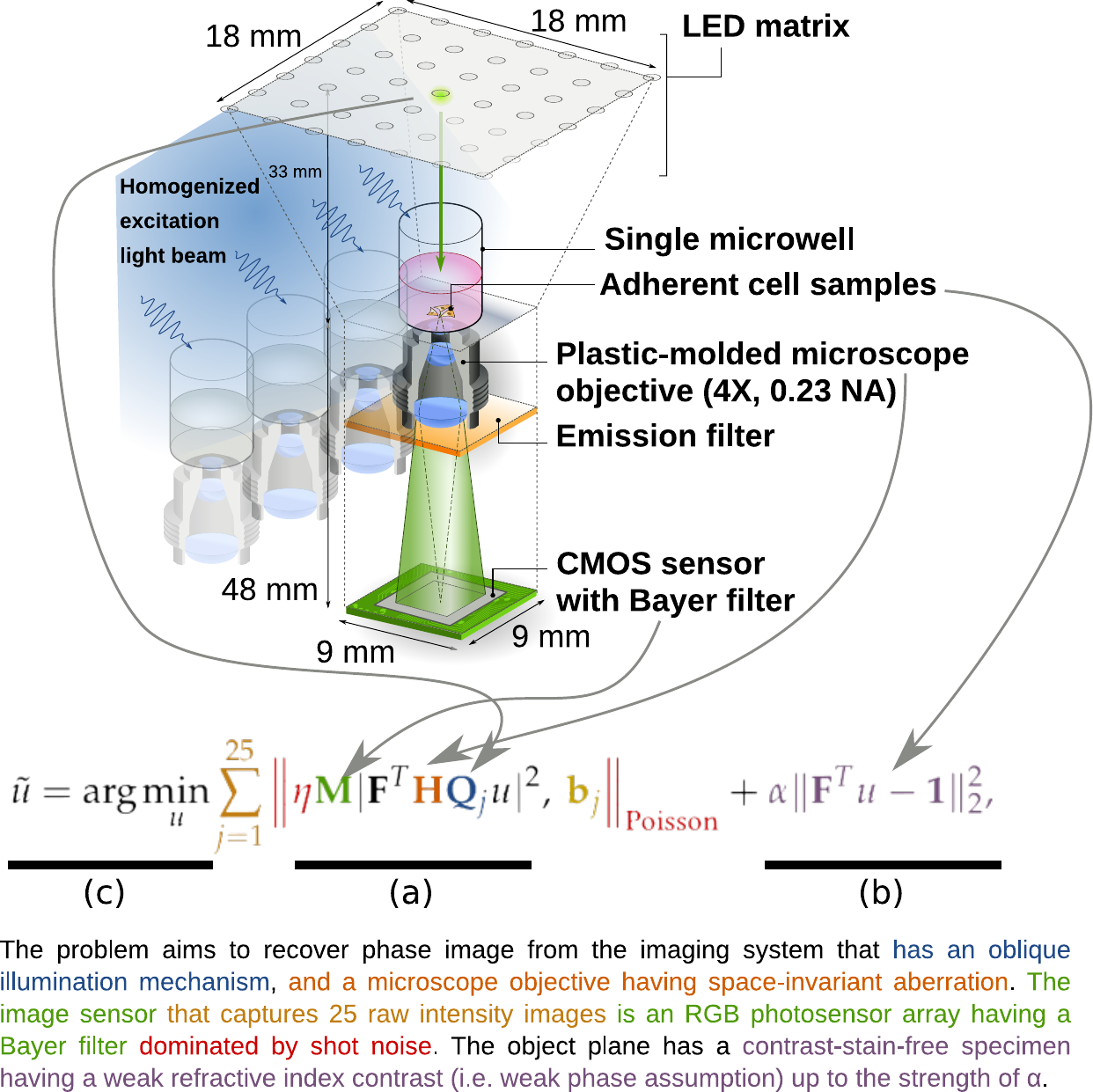}
\caption{{\bf Concurrent R\&{}D with 5GL, and its relation to the physical instrument and the biological sample.}
The inverse problem statement is rich in information because it describes
(a)~\emph{how} the signal is distorted as the light propagates through the imaging optics under oblique illuminations and captured by the photon-counting sensor,
the \emph{a priori} knowledge contributed by illumination, imaging optics, \& sensing hardware designers;
(b)~\emph{what} kind of live cell cultures is under observation,
the \emph{a posteriori} knowledge contributed by the multidisciplinary team of life science, diffractive optics, and computer vision experts;
Most of all, the problem statement precisely determines
(c)~\emph{why} there is a need of signal recovery, a collaboration between compute hardware acceleration and algorithm experts.
For the 96-Eyes system, the goal of the signal recovery is to restore the phase channel of the high-resolution image
given the intensity-only low-resolution images corrupted by multiplicative shot noise.
\label{fig:problem-statement}}
\end{figure}

To resolve communication frictions among diffractive optics, (bio-)imaging, computer vision, algorithm, and GPU researchers, there exists a working solution in the inverse problem research community; we can describe the imaging problem in a declarative form via a \emph{problem statement}
\begin{align}
\tilde u &= \arg\min_{u} \sum_{j=1}^{25} \left\Vert
\Gain \Bayer | \Fourier^T \Blur \ObliqueIlluminate u |^2,\
\Raw
\right\Vert_\text{Poisson} +
\alpha \Vert
\Fourier^T u - \FlatField
\Vert_2^2, \label{eq:problem-statement}\\
\Restored &= \Fourier^T \tilde u
\end{align}
Again, we analyze the semantic meaning of the expressions, not the Maths.
In plain English: ``the imaging system \emph{has} an oblique illumination mechanism, and a microscope objective \emph{having} space-invariant aberration.
The sensor \emph{is} an RGB photodetector array \emph{having} a Bayer filter dominated by shot noise.
The object plane \emph{has} a contrast-stain-free specimen \emph{having} a weak refractive index contrast (i.e. weak phase assumption) up to the strength of $\alpha$.''

The link between the mathematical expression and its layman description are illustrated in Fig.~\ref{fig:problem-statement}.

Notice the \emph{declarative} language pattern: (Noun)-is-a-(Noun) or (Noun)-has-a-(Noun). It concisely describes the 96-Eyes imaging hardware, namely the illumination, imaging optics' imperfection, and the sensor noise model (aka the \emph{what}). Also, through the maximum/minimum likelihood estimation operators, one can infer the underlying the signal restoration goals (super-resolution and phase retrieval, aka the \emph{why}).
The \emph{a posteriori} knowledge about the biological cell culture under observation (i.e.~stain-free, monolayer cell culture with weak-phase assumption) is also revealed in the problem statement.

Problem~\ref{eq:problem-statement} is directly coded in the ProxImaL language\cite{Heide2016},
preserving the \emph{a priori} and \emph{a posteriori} knowledge of the 96-Eyes project:
\begin{lstlisting}[language=Python]
from proximal.lin_ops import (elemwise_multiply, bayer, crop, fft_inv)
from proximal.prox_fns import (poisson_norm, sum_squares)
from proximal import Variable
from numpy import complex64

u = Variable((512, 512), dtype=complex64)

data_fidelity = [
  poisson_norm(
    eta * bayer(fft_inv(
      elemwise_multiply(H,
        crop(offset[j],  u, width=256)
      )
    ), raw=b[j])
  for j in range(1, 25)
]

weak_phase_assumption = alpha * sum_squares(fft_inv(u) - 1)

problem_statement = data_fidelity + weak_phase_assumption
\end{lstlisting}
... and the boilerplate code is omitted in the listing for the sake of clarity.

\section{Solving problems by reframing and rewriting the problem statement}

\subsection{Problem reframing by the convex optimization domain expert}
\def\LinearConstraint{\mathbf{K}}

\begin{figure}[tb]
\centering\includegraphics[width=\columnwidth]{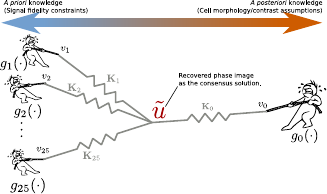}

\caption{From the perspective of convex optimization expert, FPM phase image recovery problem can be conceptualized as the consensus driven problem having multiple actors applying external constraints. The position $\tilde u$ that neutralizes all external influences is the recovered phase image.
Photo courtesy of Stanley Chan of Univ.~of Purde.
\label{fig:consensus-model}}
\end{figure}

The benefit of a declarative problem statement can be demonstrated by observing the interactions among optics, computer vision (CV), and convex optimization algorithm (CVX) researchers through the use of 5GL.
Previously, diffractive optics researchers model the 96-Eyes hardware with the wavefront propagation physics, which results in the signal fidelity term, an \emph{a priori} knowledge.
Concurrently, the CV experts work with microbiologists to derive the Maths by observing the stain-free, monolayer cell culture, contributing to the \emph{a posteriori} knowledge.
Both realms of knowledge are then integrated into the problem statement directed coded in the 5GL.
Now, the CVX researchers \emph{reframe} Problem~\ref{eq:problem-statement} into a consensus-driven problem:
\begin{subequations}
\begin{align}
\tilde u &= \arg \min_{u \in \mathbb{C} }
\sum_{j=0}^{25} g_j\left( \LinearConstraint_j u \right) \\
g_0(v) &=  \alpha \Vert v - \FlatField \Vert_2^2 &
\LinearConstraint_0 &= \Fourier^T \label{eq:weak-phase} \\
g_j (v) &= \left\Vert \Gain \Bayer  | v |^2,\
\mathbf{b}_j
\right\Vert_\text{Poisson} &
\LinearConstraint_j &= \Fourier^T \Blur \ObliqueIlluminate \label{eq:fidelity} \\
&\forall j=1,\ldots, 25. \notag
\end{align}
\end{subequations}
as illustrated in Fig.~\ref{fig:consensus-model}.
In plain English, ``the problem should be solved by achieving a consensus between:

\begin{description}
\item[Requirement~\eqref{eq:weak-phase}] the restored image (as complex wave vector field), must follow the weak phase assumption, unless the signal-to-noise ratio (SNR) exceeds a threshold proportional to $1/\alpha$; and

\item[Requirement~\eqref{eq:fidelity}] the low-resolution, aberrated signal $v_j$ must match the $j$-th raw intensity measurements through the Bayer filter array;
$v$ must minimize the multiplicative noise.''
\end{description}
Notice the change of semantic meaning in the reframed problem statement through the viewpoint of the CVX experts;
they are effectively solving a new problem even though the original design intent is preserved in the Maths
(i.e.~imaging optics specifications, illumination mechanisms, biological sample assumptions) built into the statement.

In practice, as the imaging optics and alignment specifications improves over the span of the project timeline, the original problem statement is refined progressively.
Similarly, CV experts work with life sciences and consumable teams to relax the weak-phase assumption constraints for contrast-stained cell lines, again through the use of 5GL.
As time progresses, all stakeholders are informed of the up-to-date changes at the 5GL level encouraging effective communications among domains, without resorting to deciphering the original design intent from the imperative algorithm code in 3GL.

In other words, the use of 5GL as the primary medium of communication empowers the domain experts to (re)define and analyze the problem, thus \emph{lifting} the thinking process above the (3GL) code level.

In the context of the 96-Eyes project, the problem reframing step is implemented in 5GL code as:
\begin{lstlisting}[language=Python]
from optical_needs import problem_statement, u

offsets = [
  problem_statement[j].crop_op.offset
  for j in range(1, 25)
]

new_data_fidelity = poisson_norm(
    eta * bayer(fft_inv(
      elemwise_multiply(H,
        offsets,  u, width=256)
      )
    ), raw=b)

weak_phase_assumption = problem_statement[-1]

reframed_problem = new_data_fidelity + weak_phase_assumption
\end{lstlisting}

\subsection{Problem rewriting step optimizes the algorithm without altering the original design intent}

After reframing the problem, the CVX expert performs the problem rewriting procedure, going through operator absorption, grouping, splitting, and (diagonal) scaling to ensure fast algorithm convergence. Now we have a new problem formulation:
\begin{subequations}
\begin{align}
\frac{\tilde u}{\gamma} &= \arg \min_{u \in \mathbb{C} }
f(u) + g\left( \LinearConstraint u \right) \\
f(u) &=  \alpha \left\Vert \frac{u}{\gamma} - \Fourier \FlatField \right\Vert_2^2  \\
g(v) &= \left\Vert \Gain \Bayer | \gamma v |^2, \left[ \begin{matrix}
\vdots \\
\Raw \\
\vdots \\
\end{matrix}\right]
\right\Vert_\text{Poisson} &
\LinearConstraint &= \frac{1}{\gamma^2} \Fourier^T \Blur \left[\begin{matrix}
\vdots \\
\ObliqueIlluminate \\
\vdots
\end{matrix}\right]
\end{align}
\end{subequations}
having the normalization factor $\gamma^2 > 0$ defined as the maximum singular value of the gram matrix $\LinearConstraint^T \LinearConstraint$.
Note that Problem~4 is mathematically equivalent to Problem~3: we are still describing the exact same imaging system hardware and the same underlying assumptions, e.g. noise model, unstained weak phase object.

\begin{figure}
\centering\includegraphics[width=0.8\columnwidth]{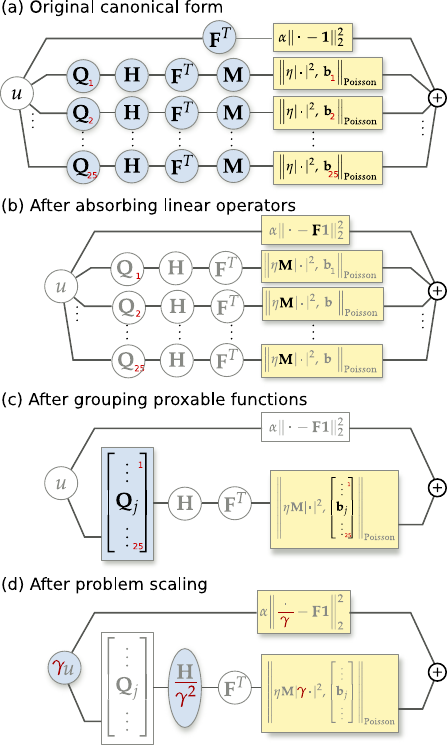}
\caption{\label{fig:proximal-working-principle}%
{\bf Problem reframing and rewriting steps performed by convex optimization (CVX) experts.}
(a)~The problem statement illustrated as a direct acyclic graph.
The graph, which encodes the sum of penalty functions with daisy-chained linear operators, undergoes
(b)~absorption,
(c)~grouping, and
(d)~scaling to improve the algorithm robustness and convergence rate.
The problem splitting step is not shown here.
These steps can be machine automated by optimizing compilers such as the ProxImaL language.\cite{Heide2016}
Terminologies:
$\Fourier^T u$: original stain-free cell image;
$\ObliqueIlluminate$: oblique illumination;
$\Blur$: optical aberration;
$\Fourier^T$: projection from the back pupil plane to the image plane;
$\Bayer$: Bayer filter of the RGB image sensor;
$\Raw$: distorted intensity images.%
}
\end{figure}

The problem (re-)formulation technique is well known by the CVX community with deterministic behavior.
Hence, the same procedure can be automated by machine algebraic transformation,
a defining feature of all modern optimizing source code compilers\cite{Lattner} (Fig.~\ref{fig:proximal-working-principle}):
\begin{lstlisting}[language=Python]
from proximal.optimizer import (absorb, group, split, scale)
from algorithm_needs import reframed_problem

optimized_problem = scale(
  split(
    group(
      absorb(
       reframed_problem
))))
\end{lstlisting}

\subsection{Reframing complex problems require bilateral accountability across domains}

The rewritten problem statement in Problem~4 underscores the value of concurrent R\&{}D workflow for effective adoption of 5GL.

In traditional waterfall-style settings, the vertically stacked operator $[\ldots\ \ObliqueIlluminate\ \ldots]^T$, presented in 5GL,
would counter the intuitions of domain expert in diffractive optics,
who originally developed these operators targeting sequential illumination behavior of the hardware.
It highlights a deeper challenge in a hardware-driven algorithm research environment:
as each oblique illumination corresponds to a single image capture step of the single CMOS image sensor,
each iteration of the reconstruction algorithm is supposed to closely track the sequential capture steps.
Here, the new problem formulation (Problem~4) implies simultaneous illumination and sensing across all 25~directions with
25~image sensors potentially mechanically interfering with each other.
The apparent violation of physical constraints may initially appear at odds with the hardware design intent.

Hence, such problem rewriting practices, while mathematically valid, can raise concerns when conducted by downstream teams without prior discussion.
Specifically, downstream teams (e.g.\ CVX and hardware-accelerated compute) may rewrite the problem to meet computational performance needs,
but without upstream consent and alignment,
these changes may be perceived as deviating from the design-locked specifications (i.e.\ Problem~\ref{eq:problem-statement}) agreed upon by both parties.

Similar case arises with the interpretation of the weak phase assumption ($\Vert \cdot - \mathbf{1} \Vert_2^2$),
originally co-developed and extensively validated by upstream CV and life science experts.
In the reformulated problem [Problem~4 and Fig.~\ref{fig:proximal-working-principle}(b)],
the downstream team implemented this prior by enforcing sparsity in the Fourier domain ($\Fourier \mathbf{1} = \delta$),
effectively treating the phase background suppression (sub-)problem
(i.e.~the proximal operator for $\Vert \cdot - \delta \Vert_2^2$) as the weighted average between the estimated phase image $u$ and a Dirac impulse function $\delta$ in Fourier domain.
While technically sound, such modifications can introduce ambiguity if not grounded in cross-validation,
especially for upstream teams prioritizing data-driven methods over theoretical derivations.

In contrast, 5GL integrated with the concurrent R\&{}D approach fosters knowledge sharing across domains.
The transparent, common description language facilitates early-stage alignment among teams,
reducing the likelihood of unintentional rivalry, and enable joint-ownership of the ever-evolving problem formulations.

\subsection{Waterfall-style development approach can lead to suboptimal collaboration output}

Now, consider a road-not-trodden scenario where the same R\&{}D process unfolds within a traditional, waterfall-style collaboration through the use of 3GL.
In this scenario, the problem statement and subsequent reformations (Problem~1~to~4) are absent altogether.
The downstream contributors, e.g.\ the GPU and algorithm researchers, typically receive a ``finalized'' implementation of the software, written in \emph{imperative} language style (often Matlab or Python).

Deciphering the underlying optics, biology and engineering needs in the 3GL code demands meticulous documentation effort by upstream teams through inline source code comments, application notes, system specification documents, and experimental validation reports,
not to mention the risk of depleting downstream team's political capital upon the formal request.
Without the supporting documents, these finalized codebases obscure the problem's original intent.
Pressured by the promise to deliver, downstream teams often resort to direct line-by-line translations into hardware-specific 3GL/2GL code (e.g. CUDA),
severely limiting their options to explore alternate problem formulations or solutions.

Although upstream teams may offer CPU-optimized code constructs (e.g. \texttt{parfor} for multithreading, or \texttt{bsxfun} for CPU vectorized instructions) to accelerate the data-driven validation during the early conceptualization stage,
these well-intentioned code additions can inadvertently complicate downstream understanding by masking the conceptual flow of the algorithm.
In practice, GPU/FPGA engineers may need to backtrack through earlier, less-refined 3GL drafts to infer the core logic,
going through a detour that could have been avoided with a shared problem formulation.

This work friction is not unique to downstream contributors.
For instance, anticipating uneven illumination and the lensing effect of the liquid meniscus,
CV experts may apply heuristic corrections (e.g.\ high-pass filtering of the raw images ahead of phase image reconstruction),
directly within the same codebase in 3GL (e.g.\ to Algorithm~\ref{alg:fpm-code}).
While intended to improve robustness, these retrofitted solutions further obscure the underlying physical signal model,
severing links to the original hardware and the governing optical diffraction laws.
The result is often a loss of upstream insight originally embedded in the domain-specific models.
Similar observations can be made from empirical gamma correction of raw images to counteract the internal nonlinear photoresponse of low-cost, consumer-grade image sensors.

Despite the structure offered by waterfall-style workflows, it can lead to fragmentation of the output 3GL code.
By severing the links among design objectives, problem formulations, and implementations,
the collaborators risk high cognitive load and communication burden, as wells as eroding trusts across domains.

\section{Hardware-accelerated compute experts maps the algorithm blueprint, in 4GL, to the corresponding features of the computer}

After the problem rewriting step, the problem undergoes \emph{algorithm synthesis}, which ``lowers'' the 5GL to an intermediate representation in 4GL.
We will witness how the hardware-accelerated compute experts and join force to solve the problem above the (3GL) code level.
Specifically, the acceleration experts inspect the 4GL output, exploiting properties of Fourier transform to optimize the hardware specific features.
Continuing our example, for the Problem~4 featuring a two-operator splitting\cite{Lions1979} and a compute-intensive linear constraint $\LinearConstraint$,
it is more favorable to synthesize the signal restoration algorithm with the hybrid Primal-Dual ascent/descent algorithm, such as the Pock-Chambolle method.
Another side note on the vocabulary: in the inverse problem community, it is common to call an algorithm synthesis technique as \emph{a method}, and the generated (Matlab/Python/C++) source code as \emph{an algorithm}.

\def\Prox{\mathbf{prox}}

\begin{algorithm}
\caption{Pock-Chambolle method to solve Problem~4.\label{alg:pock-chambolle}}
\begin{algorithmic}[1]
\State Initialize: $x, x', \bar x \in \mathbb{C}^N, z \in \mathbb{R}^M$.
\State Require: $\sigma\tau \Vert \LinearConstraint \Vert_2^2 < 1$
\State Require: $ 1 \leq \theta < 2$
\For{$k = 1 \ldots V$}
    \State $z_{k+1/2} \gets z_{k} + \sigma \LinearConstraint \bar x_k$
    \State $z_{k+1} \gets z_{k+1/2} - \sigma \Prox_{g / \sigma} (z_{k + 1/2} / \sigma)$
    \State $x_{k+1} \gets \Prox_{\tau f} (x_k - \tau \LinearConstraint^T z_{k+1})$ \label{step:dual-update}
    \State $\bar x_{k+1} \gets x_{k+1} + \theta (x_{k+1} - x_k)$
    \If{Convergence criteria is satisfied}
    	\State Break loop.
    \EndIf
\EndFor
\State \Return{$x$}
\end{algorithmic}
\end{algorithm}

The proximal operator of $f(u)$ and $g(v)$ in Problem~4 have closed-form analytical solutions:

\begin{align}
\Prox_{\tau f} (u_0) &= \arg\min_u f(u) + \frac{1}{2\tau} \Vert u - u_0 \Vert_2^2 \notag \\
&= \frac{\Fourier \mathbf{1} \sqrt{\alpha} + u_0 / (2 \tau)}{\alpha + 1/(2\tau)} \\
\Prox_{g / \sigma} (v_0) &= \arg\min_v g(v) + \frac{\sigma}{2} \Vert v - v_0 \Vert_2^2 \notag \\
&= \left[ v_1\ v_2\ \ldots\ v_i\ \ldots \right]^T \label{eq:prox-l2-norm} \\
v_i &= e^{\angle (v_0)_i} \times \left[ \frac{|(v_0)_i| - \Gain \Bayer_{ii}  / \sigma}{2} + \right. \notag \\
&\left. \sqrt{ (\Raw)_i / \sigma + \left[ \Gain \Bayer_{ii} / \sigma - |(v_0)_i| \right]^2 / 4 } \right].
\label{eq:prox-poisson-norm}
\end{align}

To qualify as 4GL, CVX algorithm experts need to work with hardware acceleration experts by co-authoring Algorithm~\ref{alg:pock-chambolle} in high level, 4GL code,
lifting the thinking process from the imperative code.
To ensure effective communications among them, an ideal 4GL for image processing algorithm must possess three distinguishing characteristics:
\begin{description}
\item[Computer-aided design (CAD)]
4GL must describe the imaging algorithm flow in \emph{spatial} domain as a standalone CAD file;

\item[Computer-aided manufacturing (CAM)]
4GL must describe, in \emph{temporal} domain, the data flow between stack memory space, heap space, and floating-point arithmetic engines, in a standalone CAM file independent of the CAD file; and, most of all

\item[Information entanglement / Bi-directional binding]
4GL must achieve machine-aided coupling between the CAD file and the CAM file.
\end{description}

\begin{figure*}[tb]
\centering\includegraphics[width=0.75\textwidth]{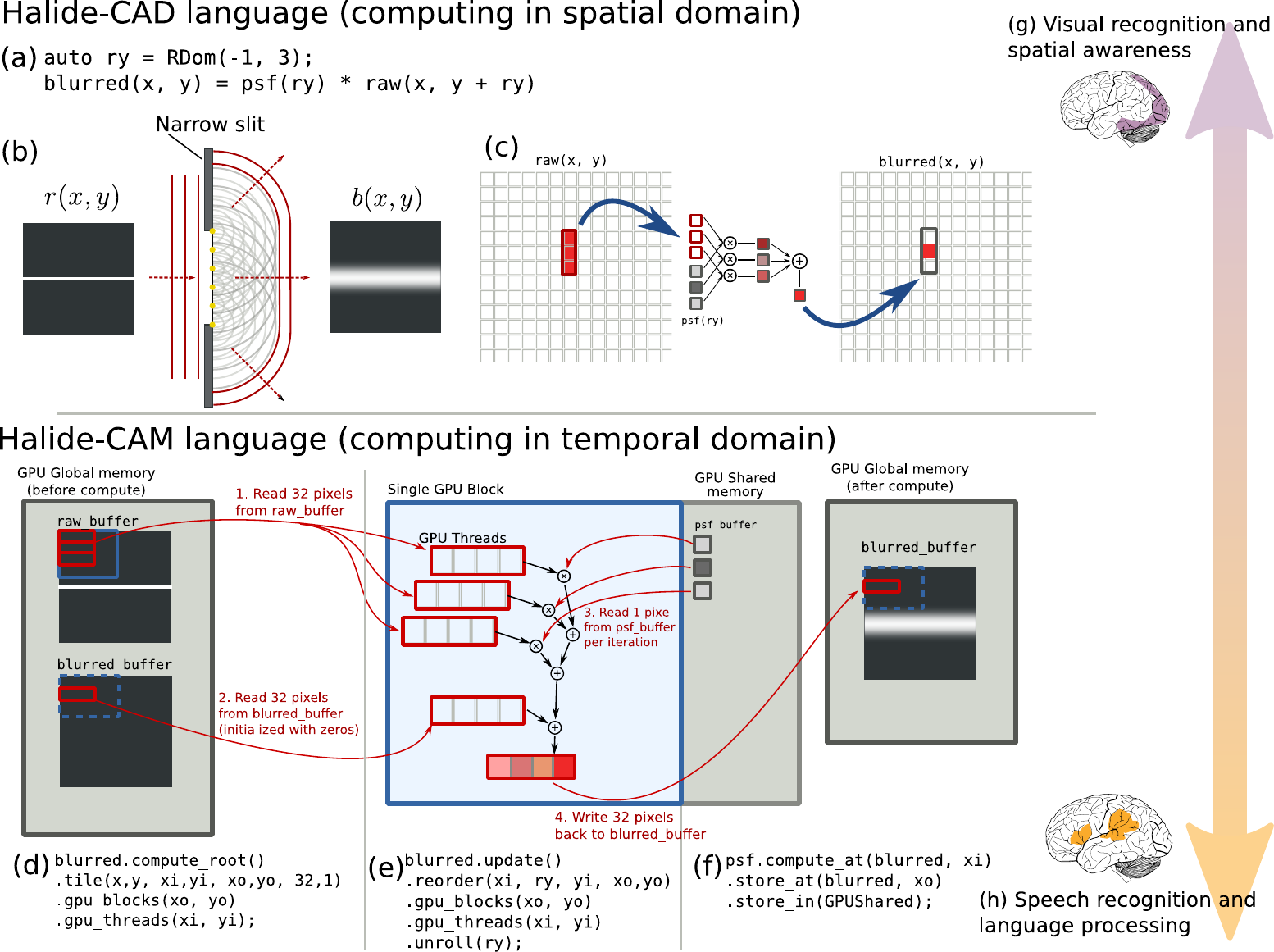}
\caption{\label{fig:halide-working-principle}%
{\bf Separation of design-versus-manufacturing concern with 4GL, illustrated in Halide language.}
(a)~Halide-CAD language aims to describe the optical blur by following the laws of Physics in the spatial domain, not the business logic of the machine.
(b)~Optical blur of an narrow slit, per Huygen's principle, can be understood as mixing of light waves from multiple point sources. The optical diffraction equation is a \emph{spatial language}.
(c)~A 3-by-3 digital blur simulates the Huygen's principle of light wave, again describing the behavior in the spatial domain.
When the digital blur algorithm (described in Halide-CAD language) is implemented on the GPU device, we compose the CAM language to describe the algorithm "flow"
among transistors, registers, and volatile memory units, in temporal domain.
(d)~Halide-CAM code to allocate the output buffer in GPU global memory, with zero-value initialization during the GPU warm-up phase;
(e)~Synthesis of the GPU-accelerated fused-multiply-add kernel;
(f)~Allocation of the GPU shared memory to cache the 3-by-3 blur kernel (one copy per GPU block).
Note that the generated GPU implementation is deliberately made suboptimal ($\geq 4$~reads from GPU global memory per pixel) for the sake of illustration.
(g)~Spatial computing involves visual recognition and spatial awareness activities, concentrated near the visual cortex region;
(h)~Temporal computing involves language processing and speech recognition activities, concentrated near the Wernickle's region.
Abbreviations: CAD -- computer-aided design; CAM -- computer-aided manufacturing, \texttt{RDom} -- reduction domain.
}
\end{figure*}

In the context of 96-Eyes project, the Halide language\cite{RaganKelley2012} (Fig.~\ref{fig:halide-working-principle}) was chosen as the primary language of the ptychographic phase image reconstruction \emph{algorithm design} and \emph{hardware/software implementation}.
The ProxImaL language described in the previous section is capable of on-demand conversion of the problem statement into Halide.

Refer to Appendix~\ref{sec:halide-explained} for how Halide enables decoupling of algorithm CAD and CAM, building solutions that survives multiple generations of GPU hardware architecture.

The ultimate strength of 4GL is revealed as the hardware acceleration experts analyzes the algorithm CAD file beyond the \emph{dual update} step, i.e. analyzing the algorithm blueprint globally.
Inspecting Step~6 and~7 of Algorithm~\ref{alg:pock-chambolle} together, one can prove that the proximal operator from Eq.~\eqref{eq:prox-poisson-norm} involves element-wise operations only.
So, one can proactively fuse the arithmetic operations from Step~6 into Step~8 to fully utilize the GPU hardware.
Traditionally, when the same algorithm were drafted in 3GL, such an algorithm ``global'' fusing strategies requires a complete rewrite of Step~6 to~8, a software bug prone process.
Here, algorithm acceleration in the global scope is realistic for the FPM algorithm in 4GL thanks to the separation of the CAD and CAM files.

To reiterate, the image processing pipeline described in 4GL, if it is ever adopted, is already considered a win over 3GL-driven development
despite the lack of connection to the design intent and cross-domain knowledge sharing offered by 5GL.

\section{5GL encourages dialogues and knowledge cross-pollinations among domain experts}

So far, we delineated the interactions among life sciences researchers and optics, CV, CVX, HW-acceleration experts as if the idea flows linearly downstream.
In the following section, I demonstrate the strength of 5GL in various FPM-related scenarios where these domains interact in self-organizing teams,
all contributing to the problem formulation statement in 5GL.
Such a concurrent research approach has eluded us in a traditional, waterfall-style research collaboration pipeline;
image restoration algorithms are often ``design-locked'' into low-level, imperative code (i.e.~3GL) before passing the knowledge to downstream domains.

\subsection{Eliminates duplicated efforts at the root}

\def\Real{\mathrm{Re}}
\def\Imag{\mathrm{Im}}

As a strawman example, researchers with the Fraunhofer-diffraction optics background, by showing goodwill,
would offer help by recognizing the wavevector field of the weak phase object as two separate image channels;
one image channel from the real component, another from the imaginary component:
\begin{equation}
u \approx ( 1 + \epsilon_\Real) + j \epsilon_\Imag,
\quad | \epsilon_\Real, \epsilon_\Imag | \ll 1,
\label{eq:separable-phase}
\end{equation}
which results in a separable, independently solvable least squared problem:
\begin{equation}
f(u) = \alpha \Vert \epsilon_\Real \Vert_2^2 +
\alpha \Vert \epsilon_\Imag \Vert_2^2,
\end{equation}
as an attempt to simplify the algorithm. In 5GL, the CVX researchers would quickly point out that the work is mathematically equivalent to Req.~\eqref{eq:weak-phase}.
Hence, all parties avoided going down a rabbit hole of data-driven validation and (GPU) acceleration in the first place.

\subsection{Adopts to new biological problems with Math rigor}

Continuing the working example, drug designers and microbiologists would point out the need to monitor cell apoptosis under influences of small molecules through the fluorescent-tagged transport protein\cite{Xu2013},
which necessities the need of quantitative phase measurements from the time-lapse image frames.
Ideally, for monolayer adherent cell lines on an optically flat substrate, the quantitative phase image should exhibit bright-colored cell structures against the black-colored uniform background in grayscale.
Ptychographic imaging experts would then raise a concern over heavy signal attenuation of quantitative phase information in the imaging hardware\cite{Tian2015,Zheng2024,Rogalski2025}.
After all, near-darkfield image is not available and non-negotiable in the 96-Eye project due to the tall chimneys of the 96-well microwell plates.

The CV expert, monitoring all the conversations in 5GL, should by now step in to incorporate the earlier approximation of Eq.~\eqref{eq:separable-phase} into
a new non-negative phase value constraint, by
\begin{align}
g_3(u) &= \mathtt{NonNegative} \left[ \mathtt{PhaseAngle} (\Fourier^T u - \FlatField) \right] \notag \\
&\approx \mathtt{NonNegative} \left[ \Imag (\Fourier^T u) \right] \notag \\
&=
\begin{cases}
0 & \epsilon_\Imag \geq 0\\
\infty & \text{otherwise},
\end{cases}
\end{align}
and then append it to the problem statement.
Now, the new problem statement is equipped with new insights for new algorithm synthesis and validation.

Drawing from prior exposure to fast-paced, waterfall-style R\&{}D workflows,
the author do note that cross-domain discussions to occur regularly,
but they often fail to translate into actionable R\&{}D tasks due to the structural limitation of imperative-style 3GL.

Consider the above hypothetical scenario in which the quantitative phase constraint is directly introduced to the FPM recovery algorithm using a traditional,
waterfall-style R\&{}D workflow with 3GL code (i.e.\ Algorithm~\ref{alg:fpm-code}).
In this case, a well-meaning upstream contributor retrofits the ``reference script'' (often in Matlab or Python) by injecting the heuristic algorithm between consecutive iterations:
separation of real and imaginary components (Eq.~\eqref{eq:separable-phase}) followed by soft-thresholding
(i.e.\ $\mathtt{ReLU}$ applied to the imaginary component.

As the need arises from data-driven validation through large-scale, historical image datasets,
the modified script was handed off to downstream teams through a series of software code patches embedded into the formal Engineering Change Order (ECO) document.
These downstream teams attempted to interpret the ECO, implement a GPU-accelerated version, and benchmark the results.
However, the effort yielded neither noticeable improvements in phase image quality nor gains in computational performance.

Simultaneously, CV researchers reviewing the 3GL code edits struggled to infer the theoretical motivation for separating the real and imaginary components.
The pixel level manipulation in Eq.~\eqref{eq:separable-phase}, which resembles a matrix transpose in linear algebra,
appeared disconnected from the core quantitative phase recovery goals.
Without access to the original formulations or design rationale of the heuristic solutions,
no teams could confidently interpret the modification as enforcing a non-negative phase constraint.

Only after downstream contributors proactively requested the relevant documentations and earlier prototypes did the underlying design intent become clear.
This detour, reliant on backward inference in 3GL rather than shared formulation in 5GL, highlighted the fragile design traceability in waterfall-style R\&{}D environment.

In contrast, the declarative language of 5GL enables all contributors to engage in joint design thinking activities above the (imperative) code level, ensuring early-alignment across domains.

\subsection{Improves the pace of bleeding-edge algorithms adoption}

\def\Eye{\mathbf{I}}

5GL encourages advanced computer vision techniques in problem formulation.
In the context of FPM,
the very first demonstration of the invention involves
a high-contrast intensity-only sample (i.e. $u\in \mathbb{R}$) imaged
through a research-grade microscope objective (i.e. $\Blur \equiv \Eye$) and
digitized by a scientific-grade monochrome camera (i.e. $\Bayer \equiv \Eye$).
In practice, it is common to capture the brightfield images in photocurrent mode (i.e. at lower camera analog gain),
and darkfield images at photon counting mode (i.e. at higher analog gain).
In 5GL, the optical design experts and CMOS image sensor experts could have described the corresponding noise models by
\begin{align}
\tilde u &= \arg\min_{u \in \mathbb{R}} G(u) + P(u) + E(u) \\
G(u) &= \sum_{j=1}^{25} \left\Vert
\Fourier^T \ObliqueIlluminate \Fourier u - \sqrt{\Raw}
\right\Vert_2^2
\tag{Thermal noise suppression from brightfield} \\
P(u) &=\sum_{j=26}^{49} \left\Vert
\Gain | \Fourier^T \ObliqueIlluminate \Fourier u |^2,\ \Raw
\right\Vert_\text{Poisson,}
\tag{Shot noise suppression from darkfield}
\end{align}
coupled with the \emph{a posteriori} knowledge provided by the computer vision expert:
\begin{align}
E(u) &= \beta \Vert \nabla u \Vert_{2,1} +
(1 - \beta) \Vert \nabla u \Vert_2^2
\tag{Edge preservation prior for high-contrast objects}
\end{align}

Upon receiving the problem statement in 5GL, it is straightforward for the CVX experts to analyze and re-write the problem through automated machine reasoning.
To be specific, they work \emph{with} 5GL optimizing compiler to absorb, group, and split the problem into the following optimized form:

\def\BrightfieldSqrt{\left[\begin{matrix}
\sqrt{\mathbf{b}_1} \\
\sqrt{\mathbf{b}_2} \\
\vdots \\
\sqrt{\mathbf{b}_{25}} \\
\end{matrix}\right]}

\def\Q{\mathbf{Q}}
\def\ObliqueIlluminateStack{\left[\begin{matrix}
\Q_1 \\
\Q_2 \\
\vdots \\
\Q_{25} \\
\end{matrix}\right]}

\begin{subequations}
\begin{align}
\hat u &= \arg \min_{u \in \mathbb{R} }
f(u) + \sum_{\ell=1}^{2} g_\ell \left( \mathbf{K}_\ell u \right)
\\
f(u) &= \left\Vert
\Fourier^T \ObliqueIlluminateStack \Fourier u - \BrightfieldSqrt
\right\Vert_2^2 \label{eq:fft-convolve} \\
g_1(v) &= \left\Vert \Gain | v |^2, \left[ \begin{matrix}
\vdots \\
\Raw \\
\vdots
\end{matrix}\right]
\right\Vert_\text{Poisson} &
\LinearConstraint_1 &= \Fourier^T \left[\begin{matrix}
\vdots \\
\ObliqueIlluminate \\
\vdots
\end{matrix}\right] \Fourier \notag \\
& \forall j=26,\ldots,49 \\
g_2(v) &= \beta \Vert v \Vert_{2,1} +
(1 - \beta) \Vert v \Vert_2^2
& \LinearConstraint_2 &= \nabla,
\end{align}
\end{subequations}
where the proximal operator for Requirement~\eqref{eq:fft-convolve} has an analytical, closed from solution involving no more than two sets of multi-plane 2D Fourier transforms per operation.
This significantly reduces the number of dual variables in the algorithms, as well as reducing the memory overhead of the hardware-accelerated compute.

\subsection{Early design feedback from HW-acceleration experts}

It is also worth to note that 5GL encourages early design feedback from hardware-acceleration experts, even though it may be considered too early to cast the problem from 5GL to 4GL.
Continuing the above example, the computer vision researcher would work with HW-acceleration experts to offer an approximation of the \emph{image gradient} operator $\nabla \approx \Fourier^T \mathrm{Diag}[D] \Fourier$.
By assuming a circular boundary condition, the approximated form has a diagonal matrix in the Fourier domain.
Such an approximation is valuable for CVX experts because they can synthesize an algorithm with alternating-direction of multiplier method (ADMM),
in which the least square optimization step of ADMM can be solved with a direct, fast Fourier transform (FFT) method\cite{Heide2016}:
\begin{align}
u^{k+1} &= \arg\min_u f(u) +
\left\Vert
\left[\begin{matrix}
\LinearConstraint_1 \\
\LinearConstraint_2
\end{matrix}\right] u - z^k + \lambda^k
\right\Vert_2^2 \notag \\
&= \frac{\Fourier^T \Q^T \Fourier \sqrt{\mathbf{b}} + (\LinearConstraint_1^T + \LinearConstraint_2^T) (z^k - \lambda^k)}
{\Fourier^T \Q^T\Q \Fourier + \LinearConstraint_1^T \LinearConstraint_1 + \LinearConstraint_2^T \LinearConstraint_2} \\
&= \Fourier^T \Delta_1 \left[ \Q^T \Fourier \sqrt{\mathbf{b}} + \Delta_2 \Fourier (z^k - \lambda^k) \right],
\end{align}
where $\Delta_1,\ \Delta_2$ are constant diagonal matrices.
The expressions may look daunting, but to the eye of GPU experts, it is a blessing in disguise.
All the diagonal matrices compile into an efficient, matrix-free, voxel-wise multiplications for the GPU.
Furthermore, upon recognizing the constant $\Fourier \sqrt{\mathbf{b}}$, one can compose the 4GL-CAM file to pre-compute and cache the Fourier-transformed raw images ahead of time on the GPU. Since the constant is stored in the high-speed GPU onboard memory, precious GPU resources can be saved from the redundant FFT-recompute and the duplicated disk-to-GPU data transfers.

Albeit the timely delivery of the 96-Eyes to the life sciences users, the author had to make tough decisions and omit a few advanced form of multidisciplinary research directions. Appendix~\ref{sec:research-directions} and Table~\ref{tab:peer-to-peer} lists new technologies (that eventually became popular after the conclusion of the 96-Eyes project) that can be integrated into the 96-Eyes' problem statement in 5GL form.

\section{Related works}

In this article, the idea of interdisciplinary communication through the 5GL is demonstrated through the case study of the 96-Eyes project.
However, the 5GL idea itself is not limited to the imaging system design niche.
Here, we briefly explore how other domains shares the vision of concurrent R\&{}D as well as problem reframing methods.

\paragraph{Model-based system engineering} System engineers calls for a high-level, joint software/hardware system model to capture design requirements.
Documentations in the form of PowerPoint slides, PDF pages, and software code does not apply because the text/code impedes communications crossing domain silos\cite{Henderson2023}.

\paragraph{Business administration} Known as the XY problem, business consultants called for problem-solving by spending more time asking the right question (the X problem), rather than implementing an attempted solution for the wrong question (the Y problem)\cite{Obermeyer2024}.

\paragraph{Database design and Structured Query Language (SQL)}
The field of relational database design was conceived to build a robust, high-volume the bank transaction database system.
Researchers advocated for a declarative query language that expresses \emph{what} data was required, rather than \emph{how} the end-users should navigate the links and perform data aggregation with imperative commands.\cite{Codd1970}
The decoupling is a distinguishing feature of 4GL.
The SQL parser internally rewrites the user-provided query statement, trading human readability with the data search performance, similar to 5GL.

\paragraph{Expert system and formal verification} As a form of classical artificial intelligence,
the pioneers behind expert systems advocate that a researcher's job should not be solving the problem by instructing the machine to perform tasks procedurally, like a programmer.
Rather, researchers should solve the problem by informing the machine \emph{declaratively} all the design constraints (5GL) and the Laws of Physics,
e.g. landing a rocket with zero velocity at ground level with minimal fuel consumption as literal statements\cite{Mattingley2011, Blackmore2016}.
Granted autonomy to seek a solution on its own, the machine rewrites the problem statement,
and generate the compute-efficient solution (3GL) given the constraints through a sequence of pure logical reasoning previously programmed by domain experts.

\paragraph{Optimizing C/C++ compilers}
Compared to classical C/C++ compiler which literally translates C/C++ code line-by-line into assembly code (2GL),
optimizing compilers\cite{Lattner, Johnson2017} parses the developer's 3GL code into an abstract syntax tree (AST) preserving the developer's original design intent.
The compiler then rewrites the AST to generate efficient 2GL code that looks completely foreign compared to the original 3GL code.
For example, when the user wants to compute the sum of first 10 integers and to print the result onto the console, the compiler can go through automated reasoning to derive the analytical formula for the sum of arithmetic sequence, followed by direct substitution of the constant~10 into the formula to evaluate the final result~(55).
So, the compiler can simply generate the executable representing a single-line \texttt{print} statement in 2GL [\texttt{print("5", "5")}].
All the redundant for-loops and summation operations are eliminated.
In this case, the optimizing compiler exhibits the distinguished features of a 5GL; the machine are granted autonomy to seek a better solution on its own (i.e. printing the hardcoded character stream ``55'' directly), treating the input C/C++ code (i.e. summing from 1 to 10) as if it is a system design constraint to satisfy, not the final decision to execute.

\paragraph{Separation of concerns for image processing algorithms} It has been proven that the Halide-CAM file for any image processing algorithm expressed only in the spatial domain, can be automatically generated by machine-automated resoning.
The Halide compiler conducts cost-benefit analysis of the user-provided Halide-CAD file, balancing the CPU/GPU recompute effort and the data cache workload.\cite{Mullapudi2016}
Other than the Halide language which is better known in mobile computing and CV community,
Google Tiramisu compiler\cite{Baghdadi2018} advocated the multi-layer, declarative programming approach to deploy large-scaled simulation problem onto the high-performance computing (HPC) nodes with high-speed network interconnect.
There are recent attempts to decouple the CAD and CAM mindsets with 4GL for algorithms (written in temporal domain) deployed on reconfigurable computing devices (a spatial device), such as the field-programmable gate array (FPGA)\cite{Hegarty2014, Rong2017, Li2020}.

\paragraph{Iterative algorithm step size tuning} If one can prove that the problem statement is strictly convex, one can improve the convergence rate of the
Pock-Chambolle algorithm with a dynamic step size\cite{Fu2020}. Recently, various disjoint step size tuning techniques have been formalized by the unifying model of critical damping of an resistor-inductor-capacitor network, a concept originated in electrical engineering domain\cite{Boyd2024}.
Such an cross-disciplinary collaboration exhibits problem reframing mindsets of 5GL; solving the CVX problem by observing it from the electrical engineer's viewpoint.

\paragraph{Iterative algorithms with plug-and-play image denoising filters}
The technique involves retrofitting the synthesized algorithm (e.g. Algorithm~\ref{alg:pock-chambolle}), in 3GL, with an external, non-linear image denoising filter, a technique experimentally proven to improve the FPM image quality and numerical stability\cite{Sun2019}.
Beyond the FPM community, the plug-and-play technique demonstrates global convergence of other image recovery problems given a high-quality initial guess\cite{Heide2014, Sun2020}.
There exists a theoretical proof of numerical stability with 3rd party non-linear image denoising filters\cite{Zhang2015}.

\paragraph{Generative, differentiable optics}
Generative optics refers to a hardware-level inverse problem where the optical system hardware and assembly can be derived \emph{de novo} from the image distortion problem statement, a approach well aligned to the philosophy of 5GL.
It is not to be confused with the term ``generative artificial intelligence (GenAI)'', aka 7GL, coined by the large-language model (LLM) community.
In the FPM community, it has been proven that the optical hardware parameters (e.g. illumination angles, lens numerical aperture) can be optimized by gradient ascent/descent of the partial derivative of Problem~\ref{eq:problem-statement} with respect to the optical parameters\cite{Zhang2015, Chen2025a}.
New variants of the 5GL are capable of computing the analytical partial derivative of Problem~\ref{eq:problem-statement} by applying machine automated chain-rules\cite{Lai2023}, so that the optical hardware parameters (e.g. illumination angles, lens aberrations) can be optimized.

\begin{figure}
\centering\includegraphics[width=0.9\columnwidth]{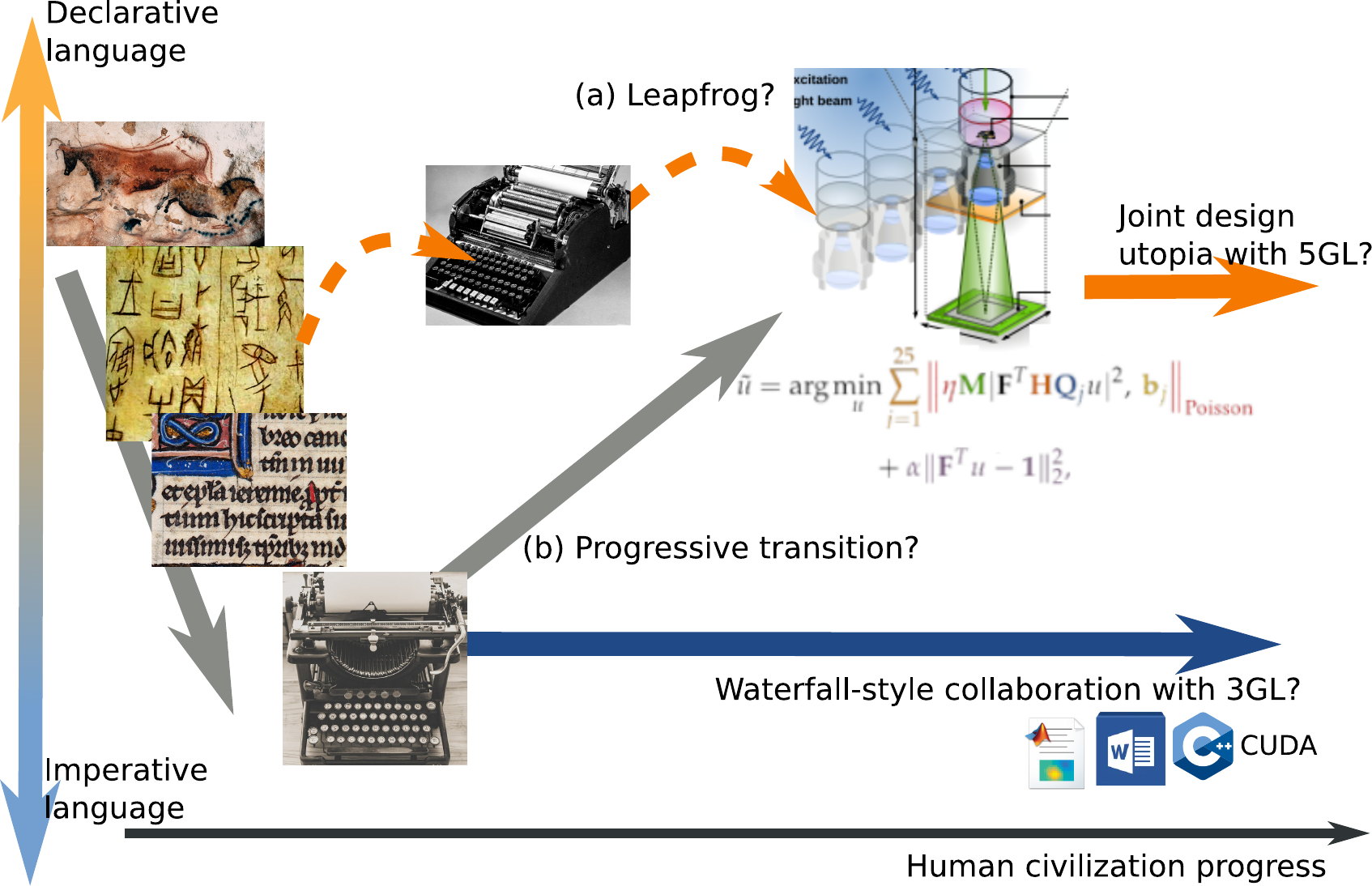}
\caption{{\bf Multidisciplinary R\&{}D in relation to human language evolution.}
The earliest collaboration takes the form of visual illustrations such as cave paintings, and picographical languages.
The invention of typewriters severed the ties between the implementations (i.e. source code) and the original design intent
(i.e. optical schematic drawings and inverse problem statement).
With the emergence of 5GL, multidisciplinary research initiatives have the potential of achieving a truly concurrent R\&{}D.
We are currently at the crossroads leading to either the joint design utopia\cite{Grady2008} (with 5GL), or preserving our current way of life with waterfall-style R\&{}D (with 3GL).
Insets, counter-clockwise from top-left:
Lascaux cave painting; bone oracle script; medieval handwritten bible; English typewriter; MingKuai typewriter invented by Yutang Lin.
\label{fig:design-utopia}}
\end{figure}

\section{The conclusions: Barriers to adoption of bilateral, high-level design thinking in collaborative R\&{}D}

At first glance, it may seem puzzling to observe a lack of adoption of concurrent R\&{}D workflow and the use of 4GL/5GL in academia and private sector.
Given the potentials to clarify design intents and foster early-alignment among domains (Fig.~\ref{fig:design-utopia}),
why do such approach remains rare in practice?

One plausible explanation lies in the contradictions between transparency and compartmentalization.
At or above the 5GL level, the problem formulation often includes detailed representation of the system behavior, hardware parameters,
and experimental conditions.
When formulized, this mission-critical knowledge effectively encodes the company's core intellectual property.
For organizations operating in competitive environment with free flow of employees in the job market,
it is natural to be wary of any workflows building on expressive, declarative languages that could leak sensitive proprietary design logic to potential rivals.
In this context, opacity (in 3GL) becomes a feature, not a bug.

A related issue emerges in multi-faculty and/or multi-corporate research initiatives,
even with efforts to promote 5GL-inspired proposal formats\cite{NIH2025, Kedia2022} in place of traditional waterfall-oriented formats\cite{NIH2024}.
The richness of 5GL-style problem formulation and the (4GL or 3GL) code generation techniques may inadvertently incentivize
contributors to extract and publish to domain-specific journals as standalone innovations,
rather than creating higher value through combined joint-design effort.
This dynamic resembles the economic principles known as Gresham's Law:
just as coins minted with precious alloys are hoarded and melted down, driving ``good'' coins out of circulation,
high-value ideas encoded in reusable components of the problem statement may be extracted and monetized separately,
preventing the systems-level, joint design thinking practices from happening in the first place.

Concurrent R\&{}D, even when well-structured, is not immune to fundamental impedance mismatch across domains.
One specific example is the treatment of signal processing in the optical research community,
where it is often perceived as a data ``rescue'' process\cite{Gibbs2021} implying corrupted measurements despite the high-precision optics and near-perfect alignment.
In this framing, the field often favors motion-free, all-optical, all-analog solutions\cite{Yaqoob2005, Solli2007, Lin2018} (akin to an upright pendulum free of external intervention to balance),
over computation-motion-optics joint designs that necessitate structured illumination and post-hoc computational image recovery
(akin to an inverted pendulum requiring active stabilization control electronics).
As such, problem formulations and the derived signal recovery algorithm are treated not as an ``digital twin'' co-evolving with the hardware design\cite{Wang2025, Chen2025b}, but an \emph{in silico} surrogate constructs inferior to the hardware counterparts.

Similar conceptual hierarchies can be observed through the choice of terminologies adopted by other domains.
For example, in multi-omic spatial biology, the common classification of collaborating domains (computer vision, bioinformatics, and clinical histopathology) into \emph{primary, secondary, and tertiary analysis} pipelines reflects an implicit order:
molecular probes and biochemical amplification assays are often framed as primary drivers of progress\cite{Wang2018, Reed2022, Arslan2023},
while downstream progress is positioned as follow-up toolboxes\cite{Altschul1990, Poplin2018, Franke2020, Moore2021} for molecular data extraction and interpretation.
Although this framing helps structure complex workflows, it can signal a organizational hierarchy among biochemistry, image acquisition, and data reconstruction,
making it more difficult to recognize novel imaging modalities\cite{Evans2022} and algorithms\cite{Huang2021,Cleary2021} as the active players in the biotechnology joint design effort.

The biochemistry-assay-first mindset is also observed in microfluidic lab-on-a-chip commercialization efforts\cite{Chin2012},
where partial\cite{Reed2022, Illumina2018} to complete\cite{Lu2016, Fuller2022} elimination of illumination/imaging optics and photodetectors are championed by cost-sensitive end-users with victory laps,
not a retreat.

More broadly, the adoption of high-level problem formulation and rewriting practice is likely influenced by the cultural divide between academic and industrial mindsets.
Based on the author's limited exposure to both worlds,
this divide likely originates from the widening rift between the \emph{discovery-oriented} education system in academia,
and the \emph{execution-oriented} vocational training in the industry.
The industry has a growing demand of fast-paced \emph{problem solvers} who consistently delivers with the \emph{can-do} attitude (\emph{Facta non verba}).
In the meantime, the academia is undergoing a reckoning of what it means by \emph{intellectuals}:
problem reframing minds practicing \emph{self-directed, relentless inquiries} amid ever changing needs in \emph{self-organizing} teams (\emph{Eadem mutata resurgo}).
When both worlds intersect, the inquiring mind is often misunderstood as deflecting the assignment (in 3GL) under the industry-wide notion that (imperative) problems have to be solved, not managed.

The 96-Eyes project is unique in the need of solving system integration challenges with bleeding-edge technologies crossing domains, justifying the need of concurrent R\&{}D practice and the problem reframing technique with 5GL.
The success of 5GL adoption is unlikely replicated on other projects where domain-specific translational research\cite{Wang2018, Reed2022, Arslan2023, Chin2012} triumphs over multidisciplinary initiatives\cite{Wang2025, Chen2025a}.
While we patiently wait for the dawn of bio-optical-computational concurrent research utopia (likely 20 to 50 years into the future if it ever materializes), let us get busy building our ``successful failures'' track record, plugging one round hole at a time.

\section*{Acknowledgements}

I thank Dr.\ Steven Diamond for the guidance on the \texttt{ProxImaL} Intermediate Representation (ProxImaL-IR) idea enabling direct baremetal code generation from Fourier Ptychographic imaging problem formulations,
especially when the linear constraint of the split variables is rank deficient.
Credits to Dr.\ Felix Heide for sharing the compiler's architectual drawings clarifying the machine-automated problem rewriting mechanisms for Flexible image signal processing ($\mathtt{FlexISP}$) firmware design applications;
Dr.\ Stanley Chan for the tug-of-war metaphor to illustrate the consensus-driven mechanism of a typical convex optimization framework.
The link between programming language generations and human language evolutions,
as well as the algorithmic insight of a ``modern'' typewriter (i.e.\ hardware implementation of 4GL/5GL) in comparison to $\mathtt{Qwerty}$ mechanical keyboards (in 2GL/3GL),
are attributed to Dr.\ Thomas Mullaney and his book \emph{The Chinese typewriter---a history}.

Valuable industry insights and perspectives on organizational dynamics and industry practices were generously shared by several tech leaders from the
molecular biotechnology (aka Biotech),
lens-free cell culture diagonstics (aka lab-on-a-chip),
and super-resolution sensing instrument (aka Techbio) sectors,
who have chosen to remain anonymous, as the article does not directly pertain to their professional affiliations,
or implying their organizations' formal position on the matter.

\section*{About the author}

Antony C.~Chan, PhD is an optical and computational imaging specialist with experience spanning the entire high throughput sensing landscape,
ranging from high-speed microscopy (i.e.\ high frame rate, single pixel), to giga-pixel superresolution camera design (i.e.\ low frame rate, high pixel count).
He has developed methods enabling million-FPS imaging cytometers,
co-invented a 96-camera live-cell imaging hardware at Caltech,
engineered tamper-proof mobile imaging to combat news disinformation,
and accelerated next-generation DNA sequencing pipelines with FPGA/GPU capabilities.
As an independent consultant, he helps clients architect advanced imaging systems end-to-end---%
from optics and motion control to firmware, algorithms,
and scalable data pipelines---%
delivering solutions that bridge cutting-edge research with deployable products for real world applications.

\myappendix

\section{Appendix}

\subsection{Separation of design-manufacture concerns with the Halide language \label{sec:halide-explained}}

To illustrate the \emph{separation of computer-aided design-vs-manufacture (CAD-CAM) concerns} offered by Halide, we can inspect the dual update step~\ref{step:dual-update} of Algorithm~\ref{alg:pock-chambolle} in the Halide-CAD file:
\begin{lstlisting}[language=C]
using Halide::Var;
using Halide::Func;

const Var x, y;
const Var kx{"wavevector_x"};
const Var ky{"wavevector_y"}
const Var j{"j_th_illumination"};

Func PhaseRamp;
PhaseRamp(x, y) = cast<int32>(
    exp(Complex(0.0, -pi) * (x + y))
);

Func Fz;
Fz(kx, ky, j) = fft2(
    Z(x, y, j) * PhaseRamp(x, y)
)(kx, ky, j);

Func CircularAperture;
CircularAperture(kx, ky) = select(kx*kx + ky*ky <= R*R, 1.0, 0.0);

Func KtZ;
KtZ(kx, ky, j) = CircularAperture(kx, ky) * conj(H(kx, ky)) * Fz(kx, ky, j);

Func XTauKtZ;
XTauKtZ(kx, ky, j) = X(kx, ky, j) - tau * KtZ(kx, ky, j);

Func XNew;
XNew(kx, ky, kz) = phase(XTauKtZ(kx, ky, j) * Equation5(XTauKtZ(kx, ky, j));
\end{lstlisting}
where the additional \texttt{PhaseRamp} operator is a spatial-domain equivalent of the \texttt{FFTShift} operation in the Fourier-domain.
Back in the earlier GPU hardware generations, \texttt{PhaseRamp} is an efficient pixel-wise arithmetic compared to the data heavy \texttt{FFTShift}.

The strength of 4GL is revealed in the Halide-CAM file that maps the algorithm design to the target compute hardware. For instance, the hardware acceleration experts specializing in Intel's research-grade CPUs would analyze the CAD code backwards, and then fuse Line~29, 26, and~23 into one single for-loop, by:
\begin{lstlisting}[language=C]
const Var xi{"inner_for_loop"};
const Var xo{"outer_for_loop");

XNew.compute_root()
   .split(x, xi, xo, 8)
   .vectorize(xi)
   .fuse(ky, j, j)
   .parallel(j);

XTauZ.compute_inline();
KtZ.compute_inline();
\end{lstlisting}
thus maximizing the large L2/L3 memory cache and multi-threading in the CPU hardware.
Similarly, constant values \texttt{CircularAperture} and \texttt{PhaseRamp} are pre-computed in a look-up table structure that takes no more than 10\% of the CPU cache, optimizing the recompute-versus-cache tradeoff:
\begin{lstlisting}[language=C]
CircularAperture.compute_root();
PhaseRamp.compute_root();
\end{lstlisting}

In contrast, if the same Halide-CAD file is analyzed by early generations of GPU experts specializing in the Nvidia-Kepler architecture, one would craft a different Halide-CAM file as
\begin{lstlisting}[language=C]
const Var xi{"x inner_for_loop"};
const Var xo{"x outer_for_loop");
const Var yi{"y inner_for_loop"};
const Var yo{"y outer_for_loop");

XNew.compute_root()
   .tile({x, y}, {xi, yi}, {xo, yo}, {32, 8})
   .gpu_threads(xi, yi)
   .gpu_blocks(xo, yo, j);

XTauZ.compute_inline();
KtZ.compute_inline();
\end{lstlisting}
by partitioning the space into grid cells of 32-by-8-by-1 voxels.
Despite the choice of voxel count per grid cell at a fraction of the maximum allowable voxel count (1024) supported by the GPU,
the data traffic to/from the GPU's shared (L2) cache is minimized in the small grid size, resulting in a faster compute.
Similarly, GPU experts would quickly identify the cheap integer-only operations for \texttt{CircularAperture} and \texttt{PhaseRamp},
and then informs Halide to pre-compute the values \emph{on demand} at the grid cell level:
\begin{lstlisting}[language=C]
CircularAperture.compute_at(XNew, xo)
    .gpu_threads(x, y);
\end{lstlisting}
to fit the GPU's shared (L2) cache, thus minimizing the data traffic to/from the GPU onboard memory.

As the GPU foundry and architecture improves over time, new generations of GPU experts can reuse the same Halide-CAD file to map
the same algorithm design blueprint to the latest hardware acceleration features baked onto the silicon.
For instance, Nvidia-Turing architecture contains integer compute silicons that are independent of the floating-point compute, enabling concurrent, pipelined compute of the integer operations and the floating-point operations.
Hence, new generations of GPU experts can now write the Halide-CAM code as follows;
\begin{lstlisting}[language=C]
CircularAperture.compute_at(XNew, xi);
\end{lstlisting}
without any measurable compute time spent on the circular aperture mask synthesis, an integer-based computation dwarfed by the more expensive floating-point computations.

\subsection{Additional R\&{}D directions uncovered by the problem statement in 5GL form \label{sec:research-directions}}

\subsubsection{Zernike-regularized lens aberration recovery}

\def\Zernike{\mathbf{Z}}

At the time of the publication of the 96-Eye article, the concept of \emph{differentiable, generative optics} were still at the infancy.
For instance, one can modify Problem~\eqref{eq:problem-statement} to recover the lens aberration instead:
\begin{align}
\hat h &= \arg\min_{c, h} \sum_{j=1}^{25} \left\Vert
\Gain \Bayer | \Fourier^T \mathrm{Diag}(h) \ObliqueIlluminate \Fourier \hat u |^2 -
\Raw
\right\Vert_2^2 + \notag \\
&\beta \Vert
\Zernike^T c - \log h
\Vert_2^2, \label{eq:pupil-recovery}
\end{align}
The Zernike prior term $\beta \Vert
\Zernike^T c - \log h
\Vert_2^2$ attempts to fit the recovered pupil to the first 30 Zernike lens aberration modes.
The prior is necessary because the 96-Eyes system omits all darkfield and NA-matching illumination in pursuit of imaging throughput.
The crucial phase transfer information are concentrated in the darkfield NA-matching raw images, and so are lost forever.

To solve for $\hat h$, one can approximate the Problem~\eqref{eq:pupil-recovery} (in 5GL) by synthesizing the corresponding (Linearized)-ADMM algorithm in 4GL code, terminating at a constant max iteration count (e.g. 5 iterations),
and then compute the partial derivative of the algorithm absolute residual $\partial \epsilon_5 / \partial \hat h$ in analytical form.
Now, for a user-specified region of interest (ROI) in the camera field of view (FOV), we can capture a large collection of raw images near the ROI, and then perform Eulerian or Newtonian gradient descent/ascent to estimate lens aberration $\hat h$.
Collecting a large collection of raw images adjacent to the ROI is necessary for 96-Eyes because of the brightfield only input raw data.

To this end, one can argue that the underlying foundational technology, machine auto-differentiation of Math expression in 4GL, aka \texttt{Autograd}, can be used to recover the lens aberration.

As of the time of the writing, the Halide language acquired the ability of machine automated partial differentiation from any imaging processing algorithm implemented purely in spatial domain\cite{Li2018}.
Similarly, researchers have demonstrated the end-to-end \emph{differentiable, generative optics} design workflow through the use of 5GL language\cite{Lai2023}, integrated with the \texttt{Autograd} feature in the PyTorch library.

\subsubsection{Algorithm synthesis with 3-operator splitting}

A more robust reconstruction algorithm is available for problem statements expressed in the canonical form $\arg\min_u f(u) + g(\LinearConstraint u) + h(u)$
where the partial derivative $\partial h/ \partial u$ has a closed-form analytical expresssion\cite{Davis2015}.
This enables computer vision experts to feed more \emph{a posteriori} knowledge to the problem statement without losing algorithm convergence performance.
For example, one can add both the weak-phase assumption $f(u) = \alpha \Vert \Fourier^T u - \FlatField \Vert_2^2$,
 and the edge-preserving assmption $h(u) = \beta \Vert \nabla u \Vert_{2,1}$,
to the unifying problem statement.

\begin{sidewaystable*}[p]
\caption{\label{tab:peer-to-peer}%
Peer-to-peer research collaboration opportunities for 96-Eye project.
In a traditional, waterfall-style collaboration approach, only the research directions in blue colors are approved.
As the name suggests, the waterfall style ensures the approved projects to be arranged along the diagonal of the table, like a staircase.
The goal of 5GL is to treat the project as a consensus-driven design problem, uncovering more collaboration opportunities (black texts).
There exist a 6GL that can uncover the bio-opto-mechanical collaborations opportunities (gray colored texts in italic);
the details are omitted for 6GL topics are out of scope of the article.
Note: the apparent asymmetry in collaboration dynamics (e.g.\ where life sciences predominantly seeks solutions from other domains without reciprocal offerings)
arises naturally from the specific project rather than an intentional critique.}

\def\gray{\textcolor{codegray}}
\def\blue{\textcolor{blue}}

\begin{tabularx}{\textheight}{cX*{7}{>{\centering\arraybackslash}X}}
\toprule
Domain needs & \multicolumn{5}{c}{Solutions offered by domain experts}\\
\cmidrule{2-6}
 & Consumable & Optics & Computer vision (CV) & Convex optimization (CVX) & Hardware (HW) acceleration \\
\midrule
Life sciences 
& \blue{\bf Biocompatibility of the microwell surface;} Tissue culture (TC-)treated surface.
& {\bf Phase-retrieval from intensity signals;} Weak-phase assumption
& {\bf Recover adherent cell morphology;} Edge-preserving TV regularization in phase channel
& {\bf Quick access to frames from time lapse experiments;} Image reconstruction with previous frames as algorithm initial guess.
& {\bf Thermal/Noise / Fan vibration constraints of benchtop instruments;} offload memory-intensive pipelines to FPGAs.\\

Consumable 
& --
& \blue{\bf Out of focus due to 96-well plate curvature and tip-tilt variations\cite{Chan2024};} Ptychographic computational re-focusing of phase channel.
& {\bf Strong phase image background due to uneven TC-treated coating;} foreground-background separation prior in the phase channel.
& {\bf Well plate edge effect due to uneven liquid evaporation per well;} \emph{in situ} wavevector calibration via differentiable optics formulation.\\

Optics 
& \gray{\bf\it FPM wavevector calibration error due to liquid meniscus variations;} sidewall coating to achieve $90^\circ$ liquid contact angle.
& --
& \blue{\bf Errors of the oblique illumination angles due to liquid meniscus;}  meniscus curvature compensating wavevector calibration.
& \blue{\bf Reconstruction failure due to low-precision optics;} computational lens aberration recovery;\\

CV 
& \gray{\bf\it Strong phase image background due to uneven TC-treated coating;\cite{Chan2019}} develop new TC-treatment assays, use optical quality substrate.
&
& --
& {\bf TV-regularization prior is compute-intensive in spatial domain;} Approximation of the $\nabla$ operator with circular boundary condition
& {\bf Acceleration of advanced, non-linear image denoising priors;} HW-accelerated non-local means filter baked onto the GPU silicon\\

CVX 
&
& {\bf Simulating image blur with 2D convolution is compute-intensive;} derive the matrix-free operations with Fourier optics.
& {\bf Poor lens aberration estimation due to lack of darkfield images;} Zernike wavefront decomposition to improve quality.
& --
& \blue{\bf Migration of CPU-accelerated code to GPUs;} decoupling algorithm CAD and CAM with 4GL, e.g. Halide\cite{RaganKelley2012}.\\
\bottomrule
\end{tabularx}
\end{sidewaystable*}

\end{document}